\def\year{2017}\relax
\documentclass[letterpaper]{article}
\usepackage{aaai17}
\usepackage{times}
\usepackage{helvet}
\usepackage{courier}
\usepackage{color}
\usepackage{graphicx}
\usepackage{epstopdf}
\usepackage{amsmath}
\usepackage{amssymb}
\usepackage{mathtools}
\usepackage{url}
\usepackage{pbox}
\usepackage{multirow}
\makeatletter
\def\@biblabel#1{}
\makeatother
\begin{document}
\title{Gated Neural Networks for Option Pricing: Rationality by Design}
\author{
\small Yongxin Yang$^{\blacklozenge,\dagger,\ddagger}$, Yu Zheng$^{\spadesuit,\dagger}$, Timothy M. Hospedales$^\blacklozenge$\\
\small EECS, Queen Mary, University of London$^\blacklozenge$, Imperial Business School, Imperial College London$^\spadesuit$\\
\small yongxin.yang@qmul.ac.uk, t.hospedales@qmul.ac.uk, y.zheng12@imperial.ac.uk \\
}
\maketitle
\begin{abstract}
We propose a neural network approach to price EU call options that significantly outperforms some existing pricing models and comes with guarantees that its predictions are economically reasonable. To achieve this, we introduce a class of gated neural networks that automatically learn to divide-and-conquer the problem space for robust and accurate pricing. We then derive instantiations of these networks that are `rational by design' in terms of naturally encoding a valid call option surface that enforces no arbitrage principles. This integration of human insight within data-driven learning  provides significantly better generalisation in pricing performance due to the encoded inductive bias in the learning, guarantees sanity in the model's predictions, and provides econometrically useful byproduct such as risk neutral density.
\end{abstract}

\section{Introduction}

Option pricing models have long been a popular research area. From a theoretical perspective, new option pricing models provide an opportunity for academics to examine financial markets' mechanics. From a practical viewpoint, market makers desire efficient pricing models to set bid and ask prices in derivative markets. The earliest and simplest pricing model, Black--Scholes \cite{bs1973} gives a rough theoretical estimate of European option price. Since then many studies attempted to find better option pricing models by relaxing the strict assumptions in Black--Scholes. The models proposed by economists usually start from a set of economic assumptions and end up with a deterministic formula that takes as input some market signals (e.g., moneyness, time to maturity, and risk-free rate). In contrast, machine learning studies solve option pricing in a  data-driven way: as a regression problem, with similar inputs to econometric models, and real market option prices as outputs. The complicated relationship between input and output (e.g., a Black--Scholes like formula) is learned from a large amount of data rather than derived from econometric axioms. Progress in data-driven option pricing can be driven by improvements in model expressivity, as well as integrating selected econometric axioms into a data-driven model as inductive bias. In this paper we achieve excellent option pricing results by contributing on both of these lines.

Regression models trained by machine learning techniques, such as kernel machines and neural networks, generalise well to out-of-sample cases as long as the training data is sufficient. Such data-driven methods give good option price estimates \cite{malliaris1993neural}, and can even surpass formula derived from economic principles. One drawback of existing data-driven approaches is that they seek a \emph{unique} solution for all options. However, learned pricing models fail on certain options, for example, some overestimate  deep out-of-the money options \cite{bennell2004black}, or underestimate options very close to maturity \cite{DugasBBNG00}. To alleviate these issues, \cite{GradojevicGK09} proposed a `divide-and-conquer' strategy, by first grouping options into sub-categories, and building distinct pricing models for each sub-category. However, this categorisation is done by manually defined heuristics, and may not be consistent with market conditions, and their changes in time. In this paper, we propose a novel class of neural networks for option pricing. These implement a divide-and-conquer method where option grouping is automatic and learned from data rather than manual heuristics. Therefore, it can dynamically adjust both option classification and refine the per-class pricing model as the market changes with time. Experiments on S\&P 500 index options show that our approach is significantly better than others.

A limitation of all the above machine learning-based methods is that while they may fit the data well (e.g., mean square error), they do not enforce some economic principles, thus ruling out their suitability for pricing in practice. E.g., option prices have theoretical bounds, the violation of which makes investors gain risk-free profit (so-called arbitrage). This motivates another \textcolor{black}{less-studied} approach to improving data-driven option pricing: opening up black box models to integrate economic axioms as constraints into learning algorithms \cite{DugasBBNG00}. From an economic perspective, this is designing a NN to make economically meaningful predictions, and from a learning perspective it is providing domain-specific inductive bias to improve generalisation and avoid overfitting. In this paper, we derive a class of gated neural networks with stronger  economic rationality guarantees than existing work. In particular our neural price predictor is the first learning based approach to carry a valid risk neutral density function, i.e., a \emph{valid probability distribution} over the future asset price  in risk neutral probability space. The terminology \emph{risk neutral} roughly implies \emph{no arbitrage} but its rigorous definition is out of scope of this work, see \cite{jeanblanc2009mathematical} for details.

Our contribution is three-fold: (1) We propose a neural network with superior option pricing performance. (2) We evaluate our method against several baselines on a large-scale dataset: it includes $5139$ trading days and $3029327$ option contracts -- this is $\textbf{70 times}$ larger than previous studies \cite{DugasBBNG00,GradojevicGK09}. (3) Our neural network model is \emph{meaningful}  in that it enforces all the necessary requirements for an economically valid (no arbitrage) call option pricing model. This results in a \emph{valid} \textcolor{black}{risk neutral density function}, from which users can extract many metrics, e.g., variance, kurtosis and skewness, that are crucial for risk management purposes.

\section{Related Work}

\vspace{0.1cm}\noindent\textbf{Econometric Methods}\quad 
Asset pricing is a very active research area in finance and mathematical finance. The oldest and most famous model for option pricing is Black--Scholes \cite{bs1973}. The biggest criticism of this model is its incompatibility with the \emph{volatility smile} behaviour in real markets due to its constant volatility assumption. The volatility smile exists due to the fact that  real-world distributions are often fat-tailed and asymmetric. Stochastic volatility models, (e.g. \cite{Heston93aclosedform}), aim to model the above smile behaviour through allowing randomness of volatility, compensated for by introducing random volatility process \cite{Heston93aclosedform}. Another stream of research suggests including jumps which represent rare events in the underlying process to alleviate the smile problem. These models are called Levy models \cite{Merton76optionpricing,Kou2002jump,Madan98thevariance,BarndorffNielsen97,Peter02fine} and are able to generate volatility skew or smile. A comprehensive theoretical explanation of asset pricing models can be found in \cite{jeanblanc2009mathematical}. This paper tackles the skew/smile problem in a more data-driven way: it learns from market prices so that a model that fits the market prices well is expected to carry the same smile structure.

There are many methods for implementing option pricing models including: Fourier-based  \cite{Carr99optionvaluation}, Tree-based  \cite{cox79simplified}, Finite difference  \cite{Schwartz77valuation} and Monte Carlo methods \cite{Boyle77options}. In this paper, we employ the fractional FFT method \cite{Cooley65an} for our benchmark option pricing models as their characteristic functions are known.

\vspace{0.2cm}\noindent\textbf{Neural Network Methods}\quad
There is a long history of computer scientists trying to solve option pricing using neural networks \cite{malliaris1993neural}. Option pricing  can be seen a standard regression task for which there are many established methods and neural networks (rebranded \emph{deep learning}) are one of the most popular choices.

Some researchers claim that it is an advantage of neural network (NN) methods that they do not make as many assumptions as the  econometric methods. However, NNs  are not orthogonal to econometric methods. In fact, some NN methods leverage classic econometric insights.  For example, \cite{garcia2000pricing} proposed a neural option pricing model with a Black--Scholes like formula. \cite{DugasBBNG00} chose specific activation functions and positive weight parameter constraints such that their model has the second-order derivative properties required by economic axioms. These studies suggested that introducing econometric constraints produces better option pricing models compared to vanilla feed-forward NNs. A good survey of this line of work can be found in \cite{opsurvey10}.

While some NN methods have benefited from econometric insights, these methods have always tried to find a universal pricing model for all options in the market. However, it has been shown that for deep out-of-money options or those with long maturity, NN methods perform very badly \cite{bennell2004black}. This is unsurprising because NN methods usually produce a  smooth pricing surface that fails to capture these awkward and low-volume parts of the market. \cite{GradojevicGK09} tried to address this issue by categorising options based on their moneyness and time to maturity, and training independent NNs for each class of options. Their grouping of options is based on fixed manual heuristic that is suboptimal, and does not adapt to the changing market data over time. Our method is a neural network that  exploits a similar divide-and-conquer idea, however it jointly learns the inter-related problems of separating options into groups and pricing  each group. Providing this increased model expressivity challenges our previous goal of building in econometric axioms to ensure meaningful predictions, because rationality constraints are harder to enforce in this more complex model. Thus we apply significant effort to contribute both a more expressive neural learner, and stronger rationality constraints guarantees to existing work.

\section{Methodology}
\subsection{Background}
We focus on EU call options. A call option is a contract that gives the buyer the right, but not the obligation, to acquire the underlying asset (e.g., stock) at a specified price (called strike price) on a certain future date (called maturity date). For example, at time $t=0$ (today), a company's stock is worth $\$100$, and a trader pays a certain amount (option cost) to buy a call option with strike price $\$110$ and maturity date $T=5$ (five days later). After five days, if the company's stock price is $\$120$, he can exercise the option to get the stock by paying the strike price $\$110$. If he sells the stock immediately at $\$120$, he profits $\$10$ (ignoring the option cost). If the company's stock price is  below $\$110$, the trader will not exercise the option and he has only lost the money paid to buy the option contract. The option pricing problem is: what should the price for this option at time $t=0$ be?

We denote the true option price from the market as $c$, and the estimate of the option pricing model as $\hat{c}$. The strike price ($\$110$ in above example) is $K$, and the time-to-maturity is $\tau$ ($\tau=T-t$). The underlying asset price at time $t$ is $S_t$ and the underlying asset price at time $T$ is $S_T$ ($\$120$ in above example -- but this is unknown at time $t=0$).

For a call option pricing model $\hat{c}(\cdot)$ with three inputs $K$, $S_t$, and $\tau$, with the assumption of no arbitrage we have,

\begin{equation*}
\hat{c}(K,S_t, \tau) = \mathrm{e}^{-r\tau} \int_0^\infty \max(0, S_T-K) f(S_T|S_t, \tau) \mathrm{d}S_T.
\label{opmodel1}
\end{equation*}

\noindent Here $r$ is the risk-free rate constant and $\mathrm{e}^{-r\tau}$ serves as a discount term. $f(x|S_t, \tau)$ is the conditional risk neutral probability density function for the asset price at time $T$ (i.e., $S_T$), given its price at time $t$ (i.e., $S_t$) and time difference (i.e., $\tau=T-t$). This equation can be explained intuitively: $\max(0, S_T-K)$ is the potential revenue of having this option at time $T$, and $f(S_T|S_t, \tau)$ is the probability density of that revenue, thus the integral term is in fact the expected revenue at time $T$ given the current status ($S_t$ and $\tau$) in risk neutral probability space. Because of the no arbitrage assumption, this expected revenue in the future should be discounted at risk-free rate to get the price at time $t$. (Note that we do not consider the dividend for simplicity).

Because the risk-free rate and discount term are obtained independently, what the option pricing method actually models is  the integral term, denoted $\tilde{c}$,
\begin{equation*}
\tilde{c}(K,S_t, \tau) = \int_0^\infty \max(0, S_T-K) f(S_T|S_t, \tau) \mathrm{d}S_T.
\label{opmodel2}
\end{equation*}
\noindent $\tilde{c}$ can be learned from data as a regression problem, but this does not necessarily lead to a \emph{meaningful} predictive model unless $f(\cdot)$ is a valid probability density function.

\subsection{Requirements for Rational Predictions}

We next list six conditions \cite{föllmer2004stochastic}~\ref{con:01a}-\ref{con:06a} that a \emph{meaningful} option pricing model should meet.

\begin{equation}
\frac{\partial \tilde{c}}{\partial K} \leq 0
\label{con:01a}\tag{C1}
\end{equation}
\noindent $\frac{\partial \tilde{c}}{\partial K}=\int_0^K f(S_T|S_t, \tau) \mathrm{d}S_T-1$ and $\int_0^K f(S_T|S_t, \tau)\mathrm{d}S_T$ is a cumulative distribution function $\mathbb{P}(S_T\le K)$ thus its value can not be larger than one.
\begin{equation}
\frac{\partial^2 \tilde{c}}{\partial K^2} \geq 0
\label{con:02a}\tag{C2}
\end{equation}
\noindent $\frac{\partial^2 \tilde{c}}{\partial K^2} = f(S_T|S_t, \tau)$ is a probability density function so its value can not be smaller than zero.  
\begin{equation}
\frac{\partial \tilde{c}}{\partial \tau} \geq 0
\label{con:03a}\tag{C3}
\end{equation}
\noindent This is intuitive: the longer you wait (larger $\tau$), the higher chance that the underlying asset price will  eventually be greater than the strike price. Thus the price should be non-decreasing with time to maturity.
\begin{equation}
\lim\limits_{K\rightarrow\infty} \tilde{c}(K, S_t, \tau) = \tilde{c}(\infty, S_t, \tau) = 0
\label{con:04a}\tag{C4}
\end{equation}
\noindent If the strike price is infinity, the option price should be zero because the underlying asset price is always smaller than the strike price. There is no point in trading the option. 
\begin{equation}
\tilde{c}(K, S_t, 0) = \max(0, S_t - K) \quad \text{when} \quad \tau = 0
\label{con:05a}\tag{C5}
\end{equation}
\noindent When $\tau = 0$, the option is ready to execute immediately, so its price should be exactly $\max(0, S_t - K)$ since $S_t = S_T$.
\begin{equation}
\max(0, S_t - K) \leq \tilde{c}(K, S_t, \tau) \leq S_t
\label{con:06a}\tag{C6}
\end{equation}
This boundary can be easily derived from put-call parity and payoff of the call option. Call option price can not exceed the underlying price, otherwise an investor can arbitrage by buying the stock and selling the option at same time and closing all positions when the option is expired. Note that the upper bound implies that when $K=0$ we should have $\tilde{c}(0,S_t,\tau) = \int_0^\infty S_T f(S_T|S_t, \tau) \mathrm{d}S_T = \mathrm{e}^{r\tau} S_t$ (also, $\hat{c}(0,S_t,\tau)=S_t$). Some studies \cite{Roper10arbitrage} prefer the integral formula instead of the upper bound, while they are actually the same. For the lower bound, call option price must exceed $\max(0, S_t - K)$ as option has time value.

\vspace{0.1cm}\noindent\textbf{Assumptions}\quad 
In the above we have made the assumptions: (i) the first and second-order derivative of $\tilde{c}$ with respect to $K$ exist. (ii)  the first-order derivative of $\tilde{c}$ with respect to $\tau$ exists. Before we introduce our proposed option pricing model, we make the last assumption: the pricing model is rescalable w.r.t. $S_t$:
\begin{equation}
\tilde{c}(\frac{K}{S_t},1,\tau) \coloneqq \frac{\tilde{c}(K, S_t, \tau)}{S_t}
\label{opmodel3}
\end{equation}
\noindent where the fraction term $\frac{K}{S_t}$ is usually called (inverse) moneyness and denoted as $m=\frac{K}{S_t}$.

\subsection{Single Model}

The core part of our option pricing model $y(m,\tau)$ is

\begin{equation}
y(m,\tau) \equiv \tilde{c}(m,1,\tau) \coloneqq \frac{\tilde{c}(K, S_t, \tau)}{S_t}.
\label{opmodel4}
\end{equation}

\noindent It takes two inputs: moneyness $m$ and time-to-maturity $\tau$. The objective is then to minimise the difference between the true market price of the option $c$ and the estimate $\hat{c}$ produced by the pricing model, where $\hat{c}=\mathrm{e}^{-r\tau}\tilde{c}=\mathrm{e}^{-r\tau} S_t y$. 

\begin{figure}[th]
	\centering
	\includegraphics[width=0.77\linewidth]{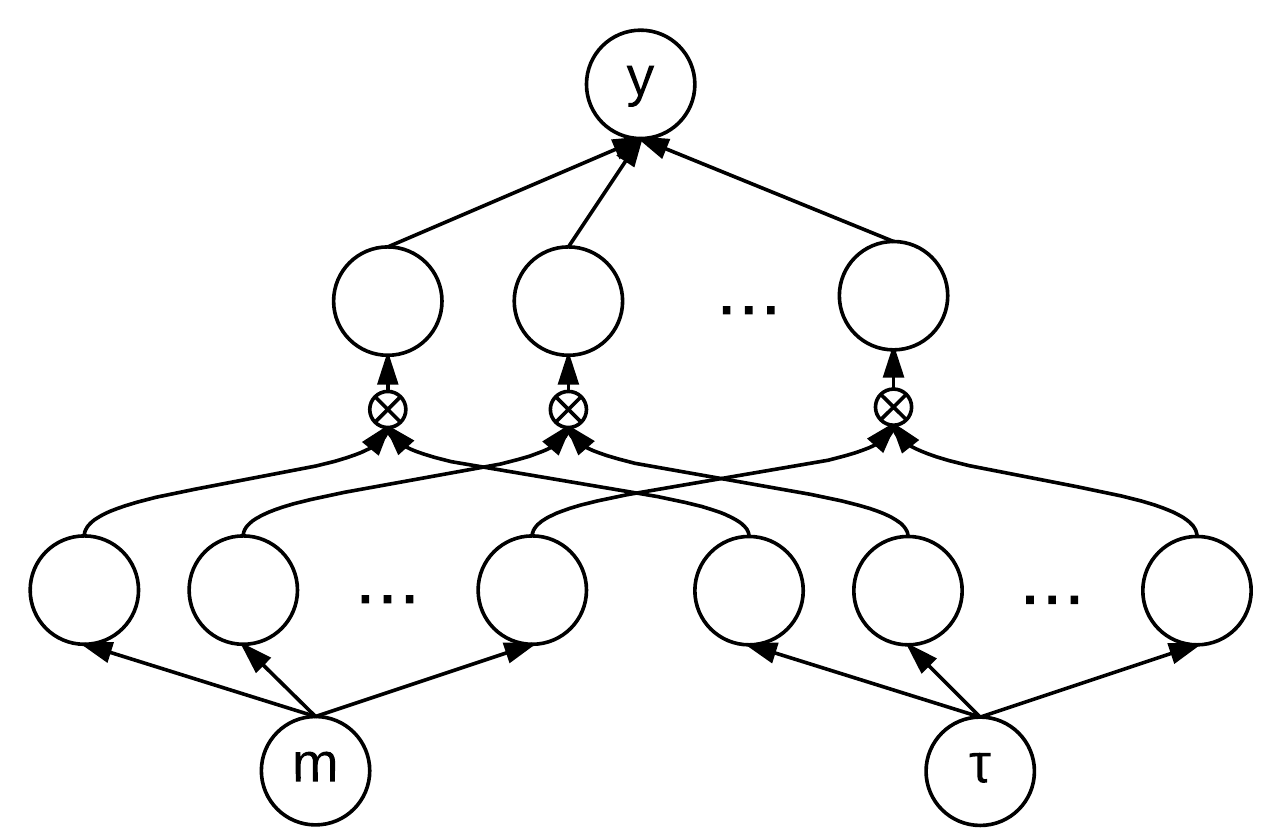}
	\caption{\small The proposed model (single). Note that bias terms exist, although they are omitted for neat appearance. $\otimes$ is the multiplication gate that outputs the product of the inputs.}
	\label{fig:singlemodel}
\end{figure}

Our pricing function $y(m,\tau)$ is modelled by a neural network  illustrated in Fig.~\ref{fig:singlemodel} and specified by  the  formula
\begin{equation}
y(m, \tau) = \sum_{j=1}^{J} \sigma_1(\tilde{b}_j-m\mathrm{e}^{\tilde{w}_j}) \sigma_2(\bar{b}_j+\tau\mathrm{e}^{\bar{w}_j})\mathrm{e}^{\hat{w}_j}.
\label{eq:singlemodel}
\end{equation}

\noindent Here $\sigma_1(x) = \log(1 + \mathrm{e}^x)$ (softplus function) and $\sigma_2(x) =\frac{1}{1 + \mathrm{e}^{-x}}$ (sigmoid function). $J$ is the number of neurons in the hidden layer. The parameters to learn are weight ($\tilde{w}$, $\bar{w}$, and $\hat{w}$) and bias ($\tilde{b}$ and $\bar{b}$) terms. We can see that this is a gated neural network \cite{Sigaud2015GatedNA} with two sides: the left-hand side takes $m$ and produces $j=1\dots J$ neurons $\sigma_1(\tilde{b}_j-m\mathrm{e}^{\tilde{w}_j})$ and the right-hand side takes $\tau$ produces $j=1\dots J$ neurons $\sigma_2(\bar{b}_j+\tau\mathrm{e}^{\bar{w}_j})$. Then  paired neurons (with the same index) from two sides are merged by a multiplication gate. Finally the $J$ penultimate layer neurons  produce the final prediction $y$ using weights $\hat{w}$.

\vspace{0.2cm}\noindent\textbf{Verifying Rationality}\quad We now show how the network of Eq.~\ref{eq:singlemodel} meets the rationality conditions laid out earlier. 
  The derivative of softplus is sigmoid function: $\sigma'_1(x) = \sigma_2(x)$, and the derivative of sigmoid is $\sigma'_2(x) = \sigma_2(x)(1-\sigma_2(x))$. Thus, we can tell that $\sigma_1(x)$, $\sigma_2(x)$, $\sigma'_1(x)$, and $\sigma'_2(x)=\sigma''_1(x)$ all produce positive values. 
Note that the weights have constrained sign: Left-branch weights are negative by imposing $-\mathrm{e}^{\tilde{w}}$, and right  and top layer weights are positive by imposing $\mathrm{e}^{\bar{w}}$ and $\mathrm{e}^{\hat{w}}$.

We can verify that Eq.~\ref{eq:singlemodel} meets conditions~\ref{con:01a}-\ref{con:03a} since

\begin{align*}
\frac{\partial y}{\partial m} = \sum_{j=1}^{J} - \mathrm{e}^{\tilde{w}_j} \sigma_2(\tilde{b}_j-m\mathrm{e}^{\tilde{w}_j}) \sigma_2(\bar{b}_j+\tau\mathrm{e}^{\bar{w}_j}) \mathrm{e}^{\hat{w}_j} \leq 0 \\
\frac{\partial^2 y}{\partial m^2} = \sum_{j=1}^{J} \mathrm{e}^{2\tilde{w}_j} \sigma'_2(\tilde{b}_j-m\mathrm{e}^{\tilde{w}_j}) \sigma_2(\bar{b}_j+\tau\mathrm{e}^{\bar{w}_j}) \mathrm{e}^{\hat{w}_j}\geq 0 \\
\frac{\partial y}{\partial \tau} =  \sum_{j=1}^{J} \mathrm{e}^{\bar{w}_j} \sigma_1(\tilde{b}_j-m\mathrm{e}^{\tilde{w}_j}) \sigma'_2(\bar{b}_j+\tau\mathrm{e}^{\bar{w}_j}) \mathrm{e}^{\hat{w}_j}\geq 0
\end{align*}

\noindent and the conditions~\ref{con:01a},~\ref{con:02a} and~\ref{con:03a} can be rewritten as,

\begin{align*}
\frac{\partial \tilde{c}}{\partial K} = \frac{\partial S_t y}{\partial K} = S_t \frac{\partial y}{\partial m} \frac{\partial m}{\partial K} = S_t \frac{\partial y}{\partial m} \frac{1}{S_t} = \frac{\partial y}{\partial m} \leq 0 \\
\frac{\partial^2 \tilde{c}}{\partial K^2} = \frac{\partial^2 y}{\partial m \partial K} = \frac{\partial^2 y}{\partial m^2} \frac{\partial m}{\partial K} = \frac{1}{S_t} \frac{\partial^2 y}{\partial m^2} \geq 0 \\
\frac{\partial \tilde{c}}{\partial \tau} =  \frac{\partial S_t y}{\partial \tau} =  S_t \frac{\partial y}{\partial \tau} \geq 0
\end{align*}

Condition~\ref{con:04a} can be easily verified as $m\rightarrow\infty$ when $K\rightarrow\infty$, and $\sigma_1(\tilde{b}_j-m\mathrm{e}^{\tilde{w}_j})=0$ when $m\rightarrow\infty$. Therefore $y=0$ and then $\tilde{c}=S_t y = 0$. This also explains why there is no bias term for the top layer.

Conditions~\ref{con:05a} and~\ref{con:06a} are hard to achieve by network architecture design (e.g., weight constraints, or activation function selection). We therefore meet them by synthesising virtual option contracts in training -- they do not exist in the real market and their true prices $c$ are equal to their theoretically estimated prices $\hat{c}$.
In detail, to meet condition ~\ref{con:05a}, we generate a number of virtual data points: For every unique $S_t$, we fix $\tau = 0$ and uniformly sample $K$ in $[0, S_t]$, and the option price should be exactly $S_t-K$. An illustration of examples of virtual options can be found in Table~\ref{tab:con5}.

\begin{table}[t]
	\centering
	\begin{tabular}{c c c c c c c}
		\hline $\tau$ & $\mathrm{e}^{-r\tau}$ & $S_t$ & $K$ & $\hat{c}$ and $c$ & $\tilde{c}$ & $y$ (expected)\\\hline
		$0$ & $1$ & $1000$ & $10$ & $990$ & $990$ & $0.9900$ \\
		$0$ & $1$ & $1000$ & $20$ & $980$ & $980$ & $0.9800$ \\
		$0$ & $1$ & $1000$ & $990$ & $10$ & $10$ & $0.0100$ \\
		.. & .. & .. & .. & .. & .. & .. \\
		$0$ & $1$ & $1100$ & $10$ & $1090$ & $1090$ & $0.9909$ \\
		$0$ & $1$ & $1100$ & $20$ & $1080$ & $1080$ & $0.9818$ \\
		$0$ & $1$ & $1100$ & $1090$ & $10$ & $10$ & $0.0091$ \\
		.. & .. & .. & .. & .. & .. & ..\\
	\end{tabular}
	\caption{\small Virtual Options for Condition~\ref{con:05a}}
	\label{tab:con5}
\end{table}

Condition~\ref{con:06a} is trickier. For the upper bound, we again synthesise virtual training options: For every unique $\tau$, we create an option with $K = 0$ corresponding to the most expensive option. Empirically the lower bound is very unlikely to be violated because (i) when $K\geq S_t$ the lower bound is $0$ -- this is met due to the neural network design (ii) when $K<S_t$, the virtual data for condition~\ref{con:05a} and the market data are highly unlikely to be mis-priced as we convert  (out-of-the-money) put options into (in-the-money) call options (details see the first part of Experiment section), so the NN model learns this lower bound from data. An illustration of examples of virtual options can be found in Table~\ref{tab:con6}.

\begin{table}[t]
	\centering
	\begin{tabular}{c c c c c c c}
		\hline $\tau$ & $\mathrm{e}^{-r\tau}$ & $S_t$ & $K$ & $\hat{c}$ and $c$ & $\tilde{c}$ & $y$ (expected)\\\hline
		$7$ & $0.98$ & $1000$ & $0$ & $1000$ & $1020$ & $1.0200$ \\
		$14$ & $0.95$ & $1100$ & $0$ & $1100$ & $1158$ & $1.0526$ \\
		.. & .. & .. & .. & .. & .. & ..\\
	\end{tabular}
	\caption{\small Virtual Options for Condition~\ref{con:06a}}
	\label{tab:con6}
\end{table}

\subsection{Multi Model}
\begin{figure}[t]
	\centering
	\includegraphics[width=0.77\linewidth]{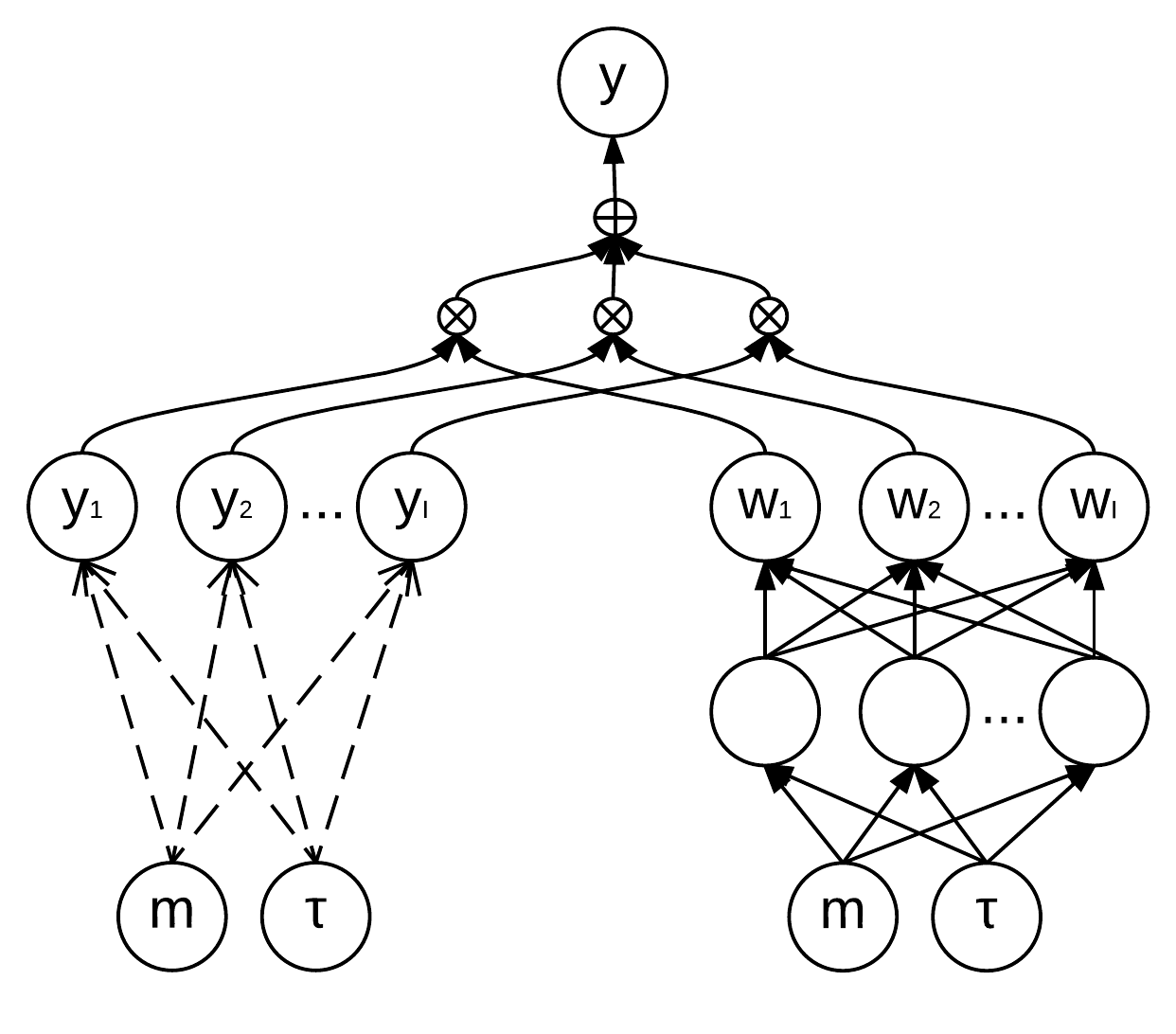}
	\caption{\small The proposed model (multi): The right side is the weight generating model, and the left side is a set of  single models. Note that the left side is not a single layer. Each $(m, \tau)\rightarrow y_i$ (linked by two dashed arrows) is realised by a full-sized single model. $\oplus$ is the addition gate that outputs the sum of the inputs.}
	\label{fig:multimodel}
\end{figure}

The previous network provides a single rational prediction model for all options. Our full model jointly trains multiple pricing models, as well as a weighting model to softly switch them.
As illustrated in Fig.~\ref{fig:multimodel}, the full model's left-hand side has $i=1\dots I$ single pricing models:
\begin{equation}
y_i(m, \tau) = \sum_{j=1}^{J} \sigma_1(\tilde{b}_j^{(i)}-m\mathrm{e}^{\tilde{w}_j^{(i)}}) \sigma_2(\bar{b}_j^{(i)}+\tau\mathrm{e}^{\bar{w}_j^{(i)}}) \mathrm{e}^{\hat{w}_j^{(i)}}
\end{equation}
Its right-hand branch is a network with one $K$ unit  hidden layer, and  the top layer has an $I$-way softmax activation function that provides a model selector for the left branch.

\begin{equation}
w_i(m, \tau) = \frac{\mathrm{e}^{\sum_{k=1}^{K}\sigma_2(m\dot{W}_{1,k} + \tau\dot{W}_{2,k} + \dot{b}_k) \ddot{W}_{k,i} + \ddot{b}_i}}{\sum_{i=1}^{I}\mathrm{e}^{\sum_{k=1}^{K}\sigma_2(m\dot{W}_{1,k} + \tau\dot{W}_{2,k} + \dot{b}_k) \ddot{W}_{k,i} + \ddot{b}_i}}
\end{equation}
Finally, the overall output $y$ is the softmax weighted average of the $I$ local option pricing models' outputs. Due to the softmax activation, the sum of weights ($w_i$'s) is one.
\begin{equation}
y(m, \tau) = \sum_{i=1}^{I} y_i(m, \tau) w_i(m, \tau).
\end{equation}
One can see the multi model as a mixture of expert ensemble \cite{jacobs1991mixExperts}, or a multi-task learning model \cite{yang15}.
The parameters of the single and multi model approaches are summarised in Table~\ref{tab:notion}.

\begin{table}[t]
\resizebox{1.00\linewidth}{!}{
\centering
\begin{tabular}{ccccc}
\hline Sym. & Shape & Comment & \pbox{1\textwidth}{Number\\ in Single} & \pbox{1\textwidth}{Number\\ in Multi} \\ 
\hline $\tilde{w}$ & $1\times J$ & Weight for moneyness & 1 & $I$\\ 
$\tilde{b}$& $J$ & Bias term for moneyness & 1 & $I$\\ 
$\bar{w}$ & $1\times J$ & Weight for time to maturity & 1 & $I$\\ 
$\bar{b}$& $J$ &  Bias term for time to maturity & 1 & $I$\\ 
$\hat{w}$ & $J\times 1$ & Weight for final pricing & 1 & $I$\\ 
\hline $\dot{W}$& $2\times K$ & Weight for input to hidden  & 0 & 1\\ 
$\dot{b}$ & $K$ & Bias term for hidden & 0 & 1\\ 
$\ddot{W}$& $K\times I$ & Weight for hidden to output & 0 & 1\\ 
$\ddot{b}$&  $I$ & Bias term for output & 0 & 1\\ 
\hline
\end{tabular}
}
\caption{\small Notation and parameter summary. Top: single pricing model. Bottom: Weighting network (right-branch) in multi model.}
\label{tab:notion}
\end{table}

\vspace{0.2cm}\noindent\textbf{Verifying Rationality}\quad It can be verified that the multi-network above still meets Conditions~\ref{con:01a},~\ref{con:03a} and~\ref{con:04a}. Conditions~\ref{con:05a},~\ref{con:06a} are again softly enforced by feed virtual data training data. The outstanding issue is that the multi-model breaks the Condition~\ref{con:02a}. To alleviate this, we use the \emph{learning from hints} trick \cite{AbuMostafa93}. 
 
Denoting the first-order derivative of $y(m,\tau)$ w.r.t. $m$ as $g(m,\tau)=\frac{\partial y}{\partial m}$, we introduce a new loss,
\begin{equation}
\sum_{p=1}^{P} \sum_{q=1}^{Q} \max(0, g(m_{p,q},\tau_q) - g(m_{p,q}+\Delta,\tau_q))
\label{eq:2ndcheck}
\end{equation}
Where $\Delta$ is an small number, e.g., $\Delta=0.001$. $Q$ is the number of unique time-to-maturity in the training set, and $P$ is the number of pseudo data generated for every unique time-to-maturity.
 Eq.~\ref{eq:2ndcheck} will push $g(m,\tau)$ to a monotonically increasing function w.r.t. $m$, thus $\frac{\partial g}{\partial m}$ (equivalently $\frac{\partial^2 y}{\partial m^2}$) tends to be larger than zero. Recall that $\frac{\partial^2 \tilde{c}}{\partial K^2} = \frac{1}{S_t} \frac{\partial^2 y}{\partial m^2}$, so Eq.~\ref{eq:2ndcheck}  fixes the negative second derivative issue. Unlike the virtual options for condition~\ref{con:05a} and~\ref{con:06a}, we do not consider the loss caused by  price difference for these data points, i.e., the data are generated for ensuring second derivative property only and we do not actually price them. 
In summary, the multi-model now also passes all the rationality checks.

\vspace{0.1cm}\noindent\textbf{Loss Functions and Optimisation}\quad
Option pricing can be sensitive to choice of loss function \cite{Christoffersen04importance}. We combine two objectives: Mean Square Error (MSE) and Mean Absolute Percentage Error (MAPE). For the multi-model, we have the extra loss in Eq.~\ref{eq:2ndcheck}. To train the NNs, we use the Adam Optimiser \cite{Kingma15adam}.

\section{Experiments}

\vspace{0.1cm}\noindent\textbf{Data and Preprocessing}\quad
The option data for S\&P500 index comes from OptionMetrics and Bloomberg, which provide historical End-of-Day bid and ask quotes. The data sample covers the period 04/01/1996-31/05/2016. The corresponding risk free rates and index dividend yields are also provided by OptionMetrics and Bloomberg.  The risk-free rates are interpolated by cubic spline to match the option maturity. Several data filters should be carried out before model calibration. Bid-ask mid-point price is calculated as a proxy for closing price. We discard in-the-money option quotes because trading is very inactive for those options thus their prices are not reliable. Furthermore, we aim to keep as many contracts as possible. We only omit  contracts with maturity less than 2 days. After these procedures, we have $3029327$ option quotes left. As our model focuses on pricing call options, we transfer put prices into call prices through put-call parity rather than discarding all put prices -- this will introduce many in-the-money call options as the complement since we discard the original in-the-money call option quotes. Time-to-maturity is annually normalised, e.g., for $\tau=7$ (seven days), the actual input is $\frac{7}{365}=0.019178$.

\begin{figure*}[t]
\centering
\includegraphics[width=1.0\linewidth]{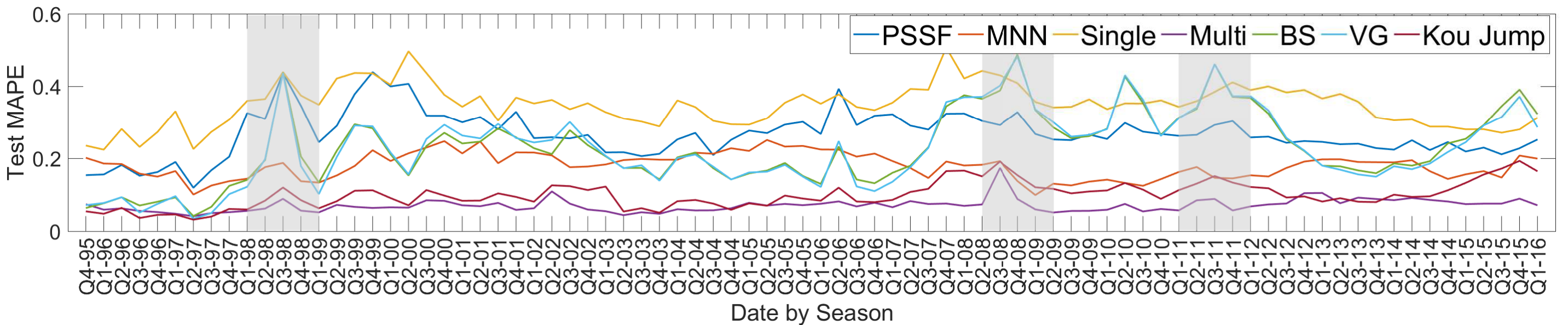}
\caption{\small Test MAPE by Seasons: The shadowed parts correspond to the following events: Dot-com bubble (1998), global financial crisis (2008), and European debt crisis (2011).}
\label{fig:result_by_season}
\end{figure*}
\subsection{Experiments I: Quantitative Comparison}

We design the experiment as follows: we train a model with five continuous trading days data, and use the following one day for testing. We compare our models denoted as \textbf{Single} and \textbf{Multi} with five baseline methods: \textbf{PSSF} \cite{DugasBBNG00}, Modular Neural Networks (\textbf{MNN}) \cite{GradojevicGK09}, Black--Scholes (\textbf{BS}) \cite{bs1973}, Variance Gamma (\textbf{VG}) \cite{Madan98thevariance}, and \textbf{Kou Jump} \cite{Kou2002jump}. For the three econometric methods\footnote{We release the code of these methods in Github: \url{https://github.com/arraystream/fftoptionlib}}, namely BS, VG, and Kou Jump, we only use the last training day's data to calibrate their parameters (see Discussion for why). For Single and PSSF, the number of hidden layer neurons is $J=5$. The number of pricing models in Multi is $I=9$ as MNN has this setting. The number of neurons in hidden layer for the right-branch weighting network of Multi is $K=5$. We report the MSE and MAPE on $(c, \mathrm{e}^{-r\tau} S_t y)$ for a meaningful comparison, though for numerical stability we train the model to (equivalently) minimise the difference on $(\mathrm{e}^{r\tau}\frac{c}{S_t}, y)$.

Both Table~\ref{tab:all_result} and Fig.~\ref{fig:result_by_season} show the superiority of our multi model in the terms of both performance and stability. We note that all methods simultaneously have drops in performance in Fig.~\ref{fig:result_by_season} at a few time points which correspond to Dot-com bubble (1998), global financial crisis (2008), and European debt crisis (2011). 

\begin{table}[t]
\centering
\resizebox{0.95\linewidth}{!}{
\centering
\begin{tabular}{c  c c  c c}
\hline
& \multicolumn{2}{c}{Train} & \multicolumn{2}{c}{Test} \\
&  MSE &  MAPE ($\%$) &  MSE &  MAPE ($\%$) \\ \hline
PSSF &  267.48 &  25.77 &  269.56 &  26.25 \\
MNN & 50.08 & 16.89 & 63.16 & 18.22 \\
Single  &  579.74 &  34.74 &  580.47 &  34.99 \\
Multi &    \textbf{9.91} &  \textbf{5.75} &   \textbf{12.11} &  \textbf{6.84}  \\\hline
BS &   63.73 &  21.64 &   64.71 &  22.42 \\
VG &   55.40 &  18.42 &   61.57 &  22.64 \\
Kou Jump &   18.37 &  8.69 &   20.13 &  9.90 \\ \hline
\end{tabular}}
\caption{\small Quantitative comparison of pricing on 3M contracts.}
\label{tab:all_result}
\end{table}

\subsection{Experiments II: Analysis of Contributions}

\begin{figure}[t]
	\centering
	\includegraphics[width=0.9\linewidth]{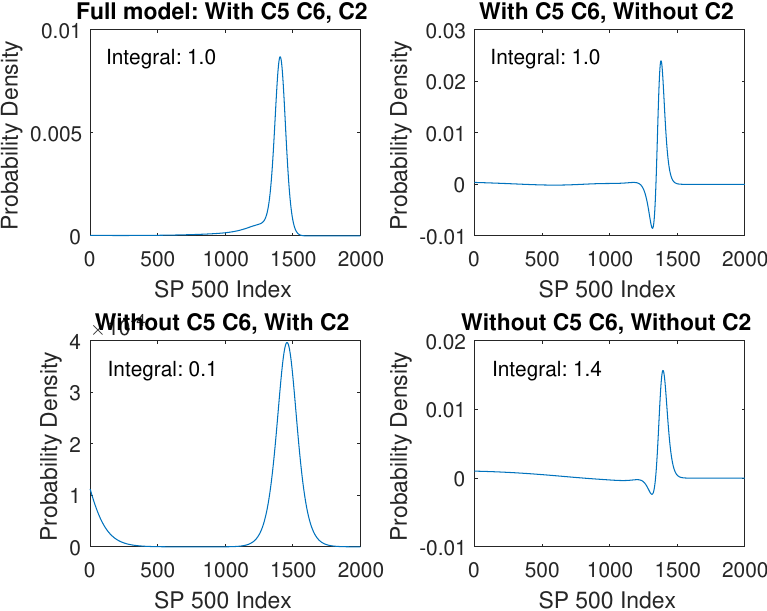}
	\caption{\small Implied distribution over future asset price. Top Left: Our multi model. Top Right: Without second derivative constraint (C2), we observe invalid negative values. Bottom Left: Without virtual options (conditions C5 and C6): we see density around zero which is senseless. Bottom Right: No derivative constraint or virtual options gives invalid and meaningless density.}
	\label{fig:exp2fig}
\end{figure}

In this section, we illustrate and validate our virtual-option strategy for meeting conditions C5 and C6, and the second derivative fix used by our multi-model for C2. We show an example when the testing day is 15th May 2008, on which the S\&P Index is $1423$. We plot the risk neutral density of the S\&P Index after $7$ days (i.e., $\tau=7$). Fig.~\ref{fig:exp2fig} shows the necessity of both virtual option contracts and positive second derivative enforcement. Both are required to generate a valid probability density, i.e., (i) non-negative and (ii) integrate to one. Furthermore, the probability density function should be economically reasonable, e.g. asset price close to zero after $\tau=7$ days should be a rare event (small probability). 

\begin{figure}[h!]
	\centering
	\includegraphics[width=1.0\linewidth]{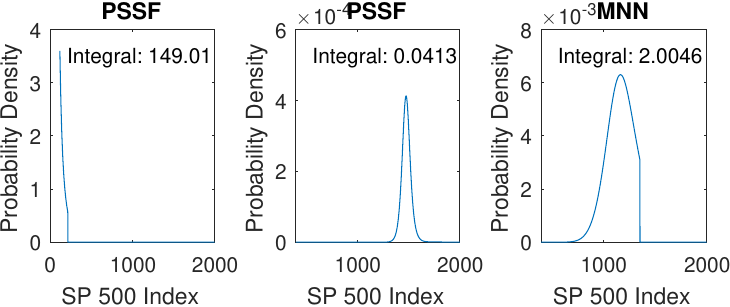}
	\caption{\small Neither PSSF nor MNN produces a valid distribution. Left: PSSF risk neural density for X-axis range $[0,2000]$. Middle: PSSF risk neural density for X-axis range $[400,2000]$ (note the difference on Y-axis scale). Right: Risk neural density of MNN.} 
	\label{fig:dugas00}
\end{figure}

Our model produces a valid density as a natural consequence of constraints C1-C6. In contrast, PSSF \cite{DugasBBNG00} in Fig.~\ref{fig:dugas00} only meets conditions C1-C3 and it produces both an invalid density and. an unreasonable large zero-price probability. Theoretically speaking, MNN \cite{GradojevicGK09} can not produce a density function because its derivative w.r.t. $K$ is not well-defined. A numerical result in Fig.~\ref{fig:dugas00} (Right) illustrates this, where we can see a discontinuous point.

\vspace{0.1cm}\noindent\textbf{Discussion}\quad We explain why we feed only one day data to the econometric methods and five days data to train the NNs. Unlike the machine learning based methods, every  parameter in econometric methods has a specific meaning. There is no analogy to increasing  model capacity through increasing the number of parameters. In contrast, NN methods offer flexible model capacity e.g. changing the number of hidden neurons. The econometric methods are designed, by principle, to fit at most one day data where $S_t$ is unique (some of them can only fit one day's data with a unique $\tau$ thus a separate step of interpolation is further required). Feeding multiple days' data to the econometric models leads to severe under-fitting and catastrophically bad performance.

In fact, requiring our model to fit multiple days' data (corresponding to multiple $S_t$ values) increases the training difficulty. In our  experiments, we found that the performance of neural network based models is \emph{negatively} related to the number of training days. The performance of NN models in Table~\ref{tab:all_result} would \emph{improve} if trained with one day data. This is against the established idea that more training data  leads to better performance. The reason is that feeding multiple days' data implicitly assumes the market structure is stable in those days. This is likely to be violated as the number of days grows, introducing a domain-shift problem. 

Why do we take this approach? Because a model that adapts to different underlying asset prices is extremely valuable when one wants to apply the option pricing model on high-frequency data: $S_t$ is no longer a constant (as underlying asset's \emph{closed} price) but a changing value (as underlying asset's \emph{current} price).  We tend to feed five days rather than one day data in our model to illustrate that it is possible to model the call option price using high frequency data. 

\section{Conclusion}

We introduced a neural network for option pricing that outperforms existing learning-based and some econometric alternatives, and comes with guarantees about the economic rationality of its outputs. In future work we will apply this option pricing model on high-frequency data, and exploit similar constraints for other finance problems such as implied volatility surface.

\vspace{0.2cm}\noindent\textbf{Acknowledgements}\quad This project received support from the European Union's Horizon 2020 research and innovation programme under grant agreement \#640891.

\end{document}